\documentclass[twocolumn,showpacs,showkeys,preprintnumbers,amsmath,amssymb,pra,floatfix]{revtex4}

\usepackage{graphicx}
\usepackage{dcolumn}
\usepackage{bm}

\begin{document}

\preprint{Proceedings of NGS-11 (June 2003, Buffalo, NY, USA), to be published in Physica E}

\title{Shake-up Processes in Intersubband Magneto-photoabsorption \\
of a Two-Dimensional Electron Gas}
\author{A.B. Dzyubenko}
\affiliation{Department of Physics, CSU at Bakersfield, Bakersfield, CA 93311, USA\\
Department of Physics, University at Buffalo, SUNY, Buffalo, NY 14260, USA}
\email{adzyubenko@csub.edu}

\date{\today}

\begin{abstract}
I theoretically study shake-up processes in photoabsorption of an interacting
low-density two-dimensional electron gas (2DEG) in magnetic fields.
Such processes, in which an incident photon creates an
electron-hole pair and simultaneously excites one electron to one of
the higher Landau levels, were observed experimentally
[D.R.~Yakovlev {\it et al.}, Phys. Rev. Lett. {\bf 79}, 3974 (1997)]
and were called {\it combined exciton-cyclotron resonance\/} (ExCR).
The recently developed theory of ExCR
[A.B.~Dzyubenko, Phys. Rev. B {\bf 64}, 241101 (2001)] allows
for a consistent treatment of the Coulomb correlations,
establishes the exact ExCR selection rules, and predicts the
high field features of ExCR. In this work, I generalize the
existing theory of high-field ExCR in the 2DEG to the case when the
{\em hole\/} is excited to higher hole Landau levels.
\end{abstract}


\maketitle

\section{\label{sec:level1} Introduction }

Correlation effects in a two-dimensional electron gas (2DEG) in
magnetic fields lead to a number of spectacular phenomena like
the Fractional Quantum Hall Effect and Wigner crystallization.
Photoluminescence (PL) spectroscopy of the 2DEG in a perpendicular
magnetic field proved to be an effective tool for studying few-
and many-body correlation effects. An interesting optical manifestation of
many-body effects in an interacting 2DEG are shake-up processes in
magneto-PL: After the recombination of the
electron-hole ($e$-$h$) pair, one electron is excited to one of the higher
Landau levels \cite{Fink96}. A closely related phenomenon, combined
exciton-cyclotron resonance (ExCR), was also identified in
low-density 2DEG systems: Here, an incident photon creates an
electron-hole pair and simultaneously excites one electron to one
of the higher Landau levels \cite{Yak97,Ossau2001}. ExCR may be
considered to be a shake-up process in magneto-photoabsorption of
the 2DEG or, a radiative Auger process as well.
These phenomena and the relation between them remain only
partially understood \cite{Fink96,Yak97,Ossau2001,PRB2001}.

In the extreme magnetic quantum limit, when electrons occupy
the lowest spin-polarized zero Landau level with filling factor
$\nu_e = 2\pi l_B^2 n_e \ll 1$, ExCR can be considered
to be a three-particle resonance $e^-_0 + {\rm photon} \rightarrow 2e$-$h$,
involving a charged system of two
electrons and one hole, $2e$-$h$, in the final state. In this limit,
it turned out possible to develop a theory of ExCR
with a consistent treatment of the Coulomb interparticle
correlations \cite{PRB2001}. Also, general and {\it exact\/}
ExCR selection rules, which follow from the existing
symmetries, magnetic translations and rotations about the magnetic
field axis, were found.  This allows one to establish the characteristic
features of the high-field ExCR; in particular, the double-peak
structure of the transitions to the first electron Landau level was
predicted. Some experimental indications of the
double-peak structure of the ExCR line were reported
recently \cite{Ossau2001}.

Magneto-PL spectroscopy proved very effective
for conduction band electrons and is able to provide a
detailed information about the rich spectrum of
electron Landau levels (LL's) and the various fine structures
associated with electron correlation effects
(see, e.g., \cite{Kukushkin} and more recent experimental  work
\cite{Yusa2001,Broocks2002a,Broocks2002} and references therein).
On the other hand, magneto-spectroscopy of valence band holes
is less effective and magneto-PL spectra of quantum well (QW)
samples containing a two-dimensional hole gas (2DHG) are
often featureless and are characterized by broad
lines, associated with the Zeeman split levels.
This is mainly because of a large difference in
electron and hole effective masses. For example,
in GaAs QW's  the electron and hole cyclotron energies,
are $\hbar\omega_{\rm ce}/B=1.7$~meV/T and $\hbar\omega_{\rm ch}/B=0.2$~meV/T, respectively, and differ
by nearly an order of magnitude \cite{GlasbergPRB2001}.
Recently, however, by examining the low energy tail of the magneto-PL spectrum of a
low-density 2DHG Glasberg {\it et al.\/} \cite{GlasbergPRB2001}
succeeded in resolving several groups of recombination
lines, each consisting of several equidistant peaks directly
related to the valence band LL's (see also \cite{Butov94,Volkov97}).
In particular, the shake-up recombination lines of the positively
charged exciton, $X^+$ were observed.

Recent progress in spectroscopy of valence band holes
in quasi-2D QW's makes it interesting to theoretically
investigate correlation effects in states in which the valence band
hole is present in higher hole LL's.
In this work, I study ExCR transitions in the 2DEG in which the hole --
in the presence of excess electrons -- are excited to higher hole LL's.
The theory predicts the shape and fine structure of the {\em high-energy tail\/} of the main
magneto-absorption peak of the 2DEG.

\section{\label{sec:theory} Theory }

I study interband transitions in the 2DEG from spin-polarized zero electron  LL $n=0\uparrow$,
$e^-_0+ {\rm photon} \rightarrow 2e$-$h$ to
the final three-particle $2e$-$h$ states  that belong to the first {\em hole\/} LL,
while electrons still reside in their zero LL's; such transitions will be called hole-ExCR (ExCR$_h$).
The classification of states according to the electron and hole
LL's is valid in sufficiently strong magnetic fields $B$ such that
\begin{equation}
      \label{highB}
 \hbar\omega_{\rm ce} \, , \,  \hbar\omega_{\rm ch} >
E_0 = \sqrt{\frac{\pi}{2}} \, \frac{e^2}{\epsilon l_B} \, .
\end{equation}
Here $l_B=(\hbar c/eB)^{1/2}$ is the magnetic length and
$E_0$ is the characteristic energy of the Coulomb interactions.
The $2e$--$h$ states can be obtained in this regime as an expansion in
electron and hole LL's.
The heavy holes are described in this work in the
effective mass approximation and the states of light holes are neglected.
The unitary transformations that allow one to construct
a complete orthonormal basis set of charged $e$-$h$ states in $B$
compatible with both axial and translational symmetries and the details of
the matrix elements calculations have been described in detail elsewhere
\cite{PRL2000,SSC,PRB2002}.
This procedure allows one to obtain the spectra of the $2e$-$h$ states
in higher hole LL's with a consistent treatment of the Coulomb
interactions.
Charged $2e$-$h$ states form families of degenerate states;
each family is labeled by the index $\nu$ that plays a role of the principal quantum number and
can be discrete (bound states) or continuous (unbound states forming
a continuum) \cite{PRL2000,SSC}.
There is a macroscopic number of degenerate states in each family labeled by the discrete
exact oscillator quantum number $k=0, 1, \ldots$.  This quantum number
is associated with magnetic translations \cite{PRL2000,SSC}
and physically describes the position of the center of the ``cyclotron'' orbit of the composite
charged complex in $B$. Each family starts with its Parent State
$ |\Psi_{k=0 \, M_z \, S_e S_h \nu}\rangle$ that has $k=0$
(rotates about the origin)
and has the largest possible in the family value of the total angular momentum projection
$M_z$; it is this value that enters into the optical selection
\cite{rmrk1}. In addition, the three-particle states are characterized by the total spin of two electrons,
either $S_e=0$ (electron spin-singlet states, $s$) or $S_e=1$ (electron spin-triplet states, $t$)
and by the spin state of the heavy hole  $S_h=3/2$,  $S_{hz}=\pm 3/2$.

\begin{figure}
\includegraphics[scale=0.5]{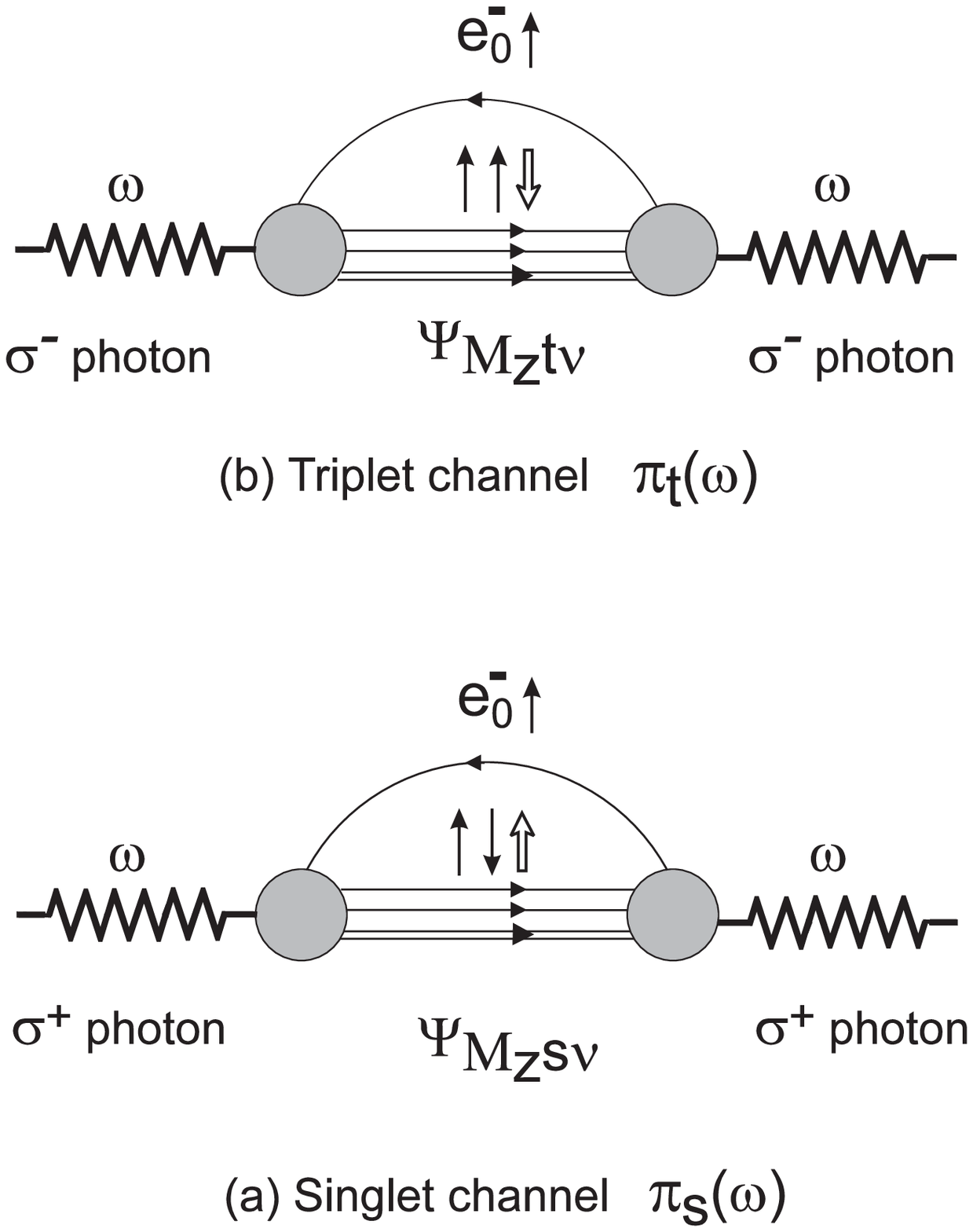}
\caption{\label{fig1}
Polarization operator $\pi(\omega)$ describing
interband absorption in the spin-polarized low-density 2DEG
in the zero LL $n=0$ with electron filling factor
$\nu_e = 2\pi n_e l_B^2 \ll 1$.
The exact selection rule is $M_z=0$; $\nu$ is the
``principal'' quantum number that labels different energy states;
see text. The absorption at frequency $\omega$ is given by
$-2{\rm Im}\,\pi(\omega)\propto n_e $.
}
\end{figure}

The exact optical selection rules for ExCR can be derived  \cite{PRB2001} as follows:
Interband transitions with $e$--$h$ pair creation
can be described by the luminescence operator
$\hat{\cal L}_{\rm PL}= p_{\rm cv} \int \! d\bm{r} \,
\hat{\Psi}^{\dagger}_{e}(\bm{r}) \hat{\Psi}^{\dagger}_{h}(\bm{r})$,
where $p_{\rm cv}$ is the interband momentum matrix element.
When an $e$--$h$ pair is photocreated in the presence
of the low-density 2DEG in the $0$-th LL,
the dipole transition matrix element can be written as
\begin{equation}
       \label{ExCR}
   D(\nu) = \langle \Psi_{M_z \, S_e S_h \nu}| \hat{\cal L}_{\rm PL}
              |e^{-}_{0} \rangle  \, ,
\end{equation}
where the final state three-particle correlations are only taken into account.
The combination of the two exact dipole selection rules, (i) conservation of the
oscillator quantum number $\Delta k = 0$ (the centers-of-rotation of charged complexes in the
initial and final states is conserved) and (ii) no change in the total angular momentum $\Delta M_z= 0$
for the {\em envelope\/} functions, leads to a very simple but
powerful result: $D({\nu}) \sim \delta_{n=0,M_z}$,
where $M_z$ is the angular momentum projection
of the Parent State in the $\nu$-th family.
Therefore, in the ExCR processes involving electrons from
the $0$-th LL, the achievable final $2e$--$h$ states must have
$M_z=0$ and may belong to {\em different} final LL's.
If the 2DEG is spin-up $\uparrow$ polarized, the
photon of $\sigma^+$ ($\sigma^-$) circular polarization produces
the electron singlet (triplet) final states (see Fig.~\ref{fig1}).

\section{\label{sec:results} Results and discussion }

The considered ExCR absorption processes involve one conduction
band electron $e_0^-\uparrow$ (the initial state) and charged three-particle $2e$–-$h$ final states
$|\Psi_{M_z s(t) \nu} \rangle$ in the electron (a) spin-singlet, $s$, and  (b) spin-triplet,
$t$, states (see Figs.~1, 2). In the states $|\Psi_{M_z s(t) \nu} \rangle$ all three
particles may be bound (thus forming charged excitons, or trions, $X^-$)
or may belong to the continuum (neutral exciton $X$ plus one electron
in a scattering state). Here and in what follows the hole spin quantum numbers
are suppressed for brevity.

\begin{figure}
\includegraphics[scale=0.53]{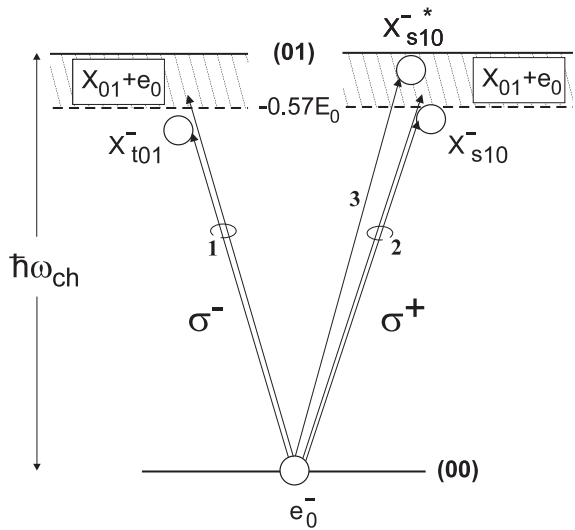}
\caption{\label{fig2} Schematic drawing of the hole-ExCR optical transitions
in the low-density 2DEG from spin-polarized $n=0$\,$\uparrow$ electron LL to
the final three-particle $2e$-$h$ states in two circular polarizations $\sigma^{\pm}$
(the hole is in its first LL).
Open dots depict three-particle bound states of charged magnetoexcitons.
The shaded area is the three-particle continuum. See also Fig.~3.
}
\end{figure}

The allowed ExCR transitions to the first hole LL and the calculated final
singlet and triplet $2e$--$h$ states  are schematically shown in Fig.~\ref{fig2}.
The shaded areas of width $0.57E_0$ in Fig.~\ref{fig2} depict the three-particle
continuum that corresponds to the motion of the bound neutral $e$-$h$ pair,
a magnetoexciton $X_{01}$ formed by the
electron $e$ in the zero LL and the hole $h$ in the first LL;
the second electron is detached from  $X_{01}$ (is in a scattering state in the zero LL\@).
The lower continuum edge lies at the $X_{01}$
ground state energy $-0.574E_0$, which, for the isolated neutral magnetoexciton,
is achieved at a {\em finite\/} center-of-mass momentum
${\bf K}$ \cite{SSC,PRB2001}.
As a result, the density of $X_{01}$ states at this energy has
the inverse square-root van Hove singularity typical for 2D
excitations.  This may lead to peculiarities in the $e^-$ scattering on
the neutral exciton $X_{01}$, in particular, one might expect formation
of quasibound three-particle states (resonances).
One such ExCR-active resonance  (see below) lying within the three-particle continuum and
denoted $X^{-*}_{s01}$ is shown in Fig.~2.
Two other open dots below the lower continuum edge in Fig.~2 represent truly bound
three-particle states, electron spin-triplet $X^-_{t01}$ and electron spin-singlet trions $X^-_{s01}$.
The binding energies of the singlet $X^-_{s01}$ and
triplet $X^-_{t01}$ states are, respectively, $0.009E_0$ and $0.024E_0$.
Note that there are many other low-lying triplet $X^-_{t01}$ and singlet $X^-_{s01}$ bound states
\cite{SSC} of negatively charged magnetoexcitons associated with the first
hole LL (not shown). The states only depicted in Fig.~2 all have $M_z=0$ and,
according to the ExCR selection rule $D({\nu}) \sim \delta_{n=0,M_z}$ \cite{PRB2001},
are ExCR-active (are the final states in the ExCR transitions).
We conclude that the ExCR transitions in 2DEG to the first hole LL can terminate both
in the bound states of charged magnetoexcitons,
$ e^-_0 + {\rm photon} \rightarrow  X^-_{s(t)01}$,  and in the continuum.
This is in contrast to the situation with transitions to the first
electron LL, when transitions to the continuum are only allowed \cite{PRB2001}.

\begin{figure}[b]
\includegraphics[angle=-90,scale=0.37]{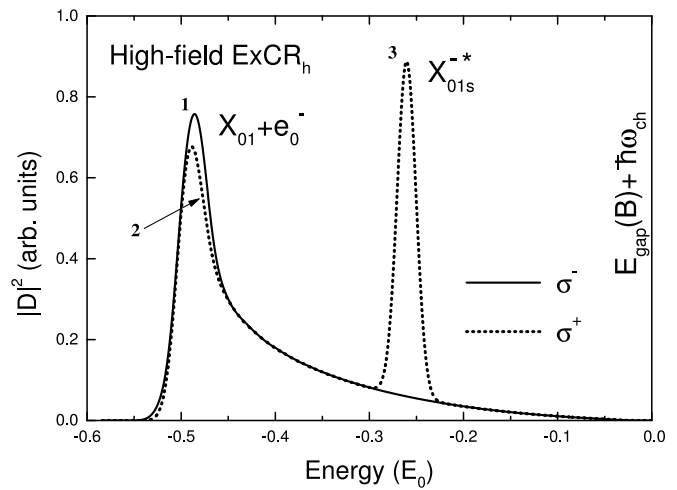}
\caption{\label{fig3} Energies and dipole matrix elements of the
hole-ExCR (ExCR$_h$) transitions from the spin-polarized $n=0$\,$\uparrow$ LL
to the first hole LL in two circular polarizations $\sigma^{\pm}$.
The representative case $E_0=\hbar\omega_{\rm ch}$ is shown.
Spectra have been convoluted with a Gaussian of the
$0.015E_0$. Labeling of the peaks is explained in Fig.~2.
The spin Zeeman energies are not included.}
\end{figure}

The calculated dipole matrix elements and energies of transitions in two
circular polarizations $\sigma^{+}$ and $\sigma^{-}$ that
terminate, respectively, in the final singlet
$| \Psi_{M_z=0 s \nu}^{(01)} \rangle$ and triplet $| \Psi_{M_z=0 t \nu}^{(01)} \rangle$
states in the zero electron and first hole LL's, $(01)$, are shown in Fig.~\ref{fig3}. These
hole-ExCR transitions require the extra photon energy $\sim \hbar\omega_{\rm ch}$
relative to the fundamental band-gap absorption $E_{\rm gap}(B)$
with final states in the lowest LL's. Also, the corresponding Zeeman energies
$\mu_{B}(g_e S_{\rm ez} + g_h S_{\rm hz})B \sim B$ must be added
to the Coulomb interaction energies $\sim B^{1/2}$ that are only shown in Fig.~3.
Because of the small binding energies of the trions $X^-_{s01}$ and $X^-_{t01}$
in the first hole LL, transitions to these states merge,  in the presence of moderate
broadening,  with the strong transitions that terminate near the lower
continuum edge (transitions~1 and 2 in Figs.~2 and 3);
these transitions dominate the hole-ExCR spectra.
The present theory predicts in high fields another
strong feature, the second higher-lying peak in the hole-ExCR that can
only be observed in the $\sigma^+$ polarization.
This polarization dependence is due to different $e$--$e$ correlations
in the final singlet ($\sigma^+$ polarization) and
triplet ($\sigma^-$ polarization) $2e$--$h$ states:
the final singlet $2e$--$h$ states are characterized by stronger
$e$--$e$ repulsion. The second ExCR peak may be associated with a formation of a quasi-bound
three-particle electron singlet state (resonance) $X^{-*}_{s01}$ within the three-particle
continuum. This is because the amplitude of finding all three particles in the same
region of real space is large for a well-defined resonance, which may lead
to large optical oscillator strengths. The existence of the $2e$--$h$
resonances is physically plausible because of the 2D van Hove
singularity in the $X_{10}$ density of states.
Another well-developed Fano-resonances were revealed recently \cite{PRB2002} in the spectra
of intraband internal $X^-$ transitions and in interband ExCR transitions to the first
electron LL \cite{PRB2001}.

The present theory shows that the main hole-ExCR peaks have intrinsic finite linewidths,
in high fields $\sim 0.1 E_0$ and have asymmetric lineshapes with high-energy tails.
It should also be noted that all ExCR transitions in 2DEG are only because of {\em electron\/} LL mixing
\cite{PRB2001} and, therefore, the hole-ExCR transitions are suppressed in strong fields as
$\nu_e |D|^2 \sim n_e l_B^2 (E_0/\hbar\omega_{\rm ce})^2 \sim B^{-2}$.
It is interesting to note here that very similar {\em final\/} states
to those in interband hole-ExCR transitions in 2DEG to higher hole LL's
may also be reached by internal {\em intraband\/} excitations \cite{PRL2002} of negatively charged
excitons $X^-$ with transition terminating in the states belonging to higher hole LL's.

It is important to note here also that while shake-up (ExCR) processes in
magneto-photoabsorption of the low-density 2DEG are allowed, the
exact selection rules \cite{PRL2000,PRB2001} prohibit shake-up processes in
magneto-PL of {\em isolated\/} negative $X^-$
and positive $X^+$ charged trions.
Common experimental observations in the dilute limit of the shake-up processes
\cite{Fink96,GlasbergPRB2001} and magneto-PL of the ``dark''
triplet $X^-$ states  (see \cite{Sanvitto} and references therein)
may be interpreted as an indication toward the relevance of the scattering by
disorder and/or the remaining 2DEG (for $X^-$) or 2DHG ($X^+$).

\section{\label{sec:summary} Summary}

I have theoretically considered combined hole-ExCR in a low-density
2DEG, a resonance in the 2DEG photoabsorption in which the hole
is excited to higher hole Landau levels.
This resonance may be observed in high magnetic fields as a fine structure
in the high-energy tail of the main magnetoabsorption peak of the 2DEG.
It has been shown, in particular, that the high-field hole-ExCR has
different absorption patterns in two different circular
polarizations $\sigma^{\pm}$, which may be useful for magneto-optical studies of
electron-hole correlations in quantum wells in strong magnetic fields.

\begin{acknowledgments}
This work is supported in part by NSF grants DMR-0203560 and DMR-0224225
and by a research grant of CSUB.
\end{acknowledgments}




\end{document}